\RequirePackage{ifpdf}
\ifpdf 
\documentclass[pdftex]{sigma}
\else
\documentclass{sigma}
\fi

\numberwithin{equation}{section}

\begin{document}

\allowdisplaybreaks

\renewcommand{\thefootnote}{$\star$}

\renewcommand{\PaperNumber}{025}

\FirstPageHeading

\ShortArticleName{Supersymmetric Quantum Mechanics and Painlev\'e IV Equation}
\ArticleName{Supersymmetric Quantum Mechanics \\ and Painlev\'e IV Equation\footnote{This
paper is a contribution to the Proceedings of the Workshop ``Supersymmetric Quantum Mechanics and Spectral Design'' (July 18--30, 2010, Benasque, Spain). The full collection
is available at
\href{http://www.emis.de/journals/SIGMA/SUSYQM2010.html}{http://www.emis.de/journals/SIGMA/SUSYQM2010.html}}}

\Author{David BERM\'UDEZ and David J. FERN\'ANDEZ C.}
\AuthorNameForHeading{D. Berm\'udez and D.J. Fern\'andez C.}

\Address{Departamento de F\'isica, Cinvestav, AP 14-740, 07000 M\'exico DF, Mexico}
\Email{\href{mailto:dbermudez@fis.cinvestav.mx}{dbermudez@fis.cinvestav.mx}, \href{mailto:david@fis.cinvestav.mx}{david@fis.cinvestav.mx}}
\URLaddress{\url{http://www.fis.cinvestav.mx/~david/}}

\ArticleDates{Received November 30, 2010, in f\/inal form March 04, 2011;  Published online March 08, 2011}

\Abstract{As it has been proven, the determination of general one-dimensional Schr\"odinger Hamiltonians having third-order dif\/ferential ladder operators requires to solve the Painlev\'e IV equation. In this work, it will be shown that some specif\/ic subsets of the higher-order supersymmetric partners of the harmonic oscillator possess third-order dif\/ferential ladder operators. This allows us to introduce a simple technique for generating solutions of the Painlev\'e IV equation. Finally, we classify these solutions into three relevant hierarchies.}

\Keywords{supersymmetric quantum mechanics; Painlev\'e equations}

\Classification{81Q60; 35G20}

\renewcommand{\thefootnote}{\arabic{footnote}}
\setcounter{footnote}{0}

\section{Introduction}

Nowadays there is a growing interest in the study of nonlinear phenomena and their corresponding description. This motivates to look for the dif\/ferent relations which can be established between a given subject and nonlinear dif\/ferential equations \cite{Sac91}. Specif\/ically, for supersymmetric quantum mechanics (SUSY QM), the standard connection involves the Riccati equation~\cite{AIS93,FFG05}, which is the simplest nonlinear f\/irst-order dif\/ferential equation, naturally associated to general Schr\"odinger eigenproblems. Moreover, for particular potentials there are other links, e.g., the SUSY partners of the free particle are connected with solutions of the KdV equation~\cite{Lam80}. Is there something similar for potentials dif\/ferent from the free particle?

In this paper we are going to explore further the established relation between the SUSY partners of the harmonic oscillator and some analytic solutions of the Painlev\'e~IV equation~($P_{\rm IV}$). This has been studied widely both in the context of dressing chains \cite{VS93,DEK94,Adl94,Spi95} and in the framework of SUSY QM \cite{ACIN00,FNN04,CFNN04,MN08}. The key point is the following: the determination of general Schr\"odinger Hamiltonians having third-order dif\/ferential ladder operators requires to f\/ind solutions of $P_{\rm IV}$. At algebraic level, this means that the corresponding systems are characterized by second-order polynomial deformations of the Heisenberg--Weyl algebra, also called polynomial Heisenberg algebras (PHA) \cite{CFNN04}. It is interesting to note that some generalized quantum mechanical systems have been as well suggested on the basis of polynomial quantum algebras, leading thus to a $q$-analogue of $P_{\rm IV}$ \cite{Spi95}.

On the other hand, if one wishes to obtain solutions of $P_{\rm IV}$, the mechanism works in the opposite way: f\/irst one looks for a system ruled by third-order dif\/ferential ladder operators; then the corresponding solutions of $P_{\rm IV}$ can be identif\/ied. It is worth to note that the f\/irst-order SUSY partners of the harmonic oscillator have associated natural dif\/ferential ladder operators of third order, so that a family of solutions of $P_{\rm IV}$ can be easily obtained through this approach. Up to our knowledge, Flaschka (1980) \cite{Fla80} was the f\/irst people who realized the connection between second-order PHA (called commutator representation in this work) and $P_{\rm IV}$ equation (see also the article of Ablowitz, Ramani and Segur (1980) \cite{ARS80}). Later, Veselov and Shabat (1993) \cite{VS93}, Dubov, Eleonsky and Kulagin (1994) \cite{DEK94} and Adler (1994) \cite{Adl94} connected both subjects with f\/irst-order SUSY QM. This relation has been further explored in the higher-order case by Andrianov, Cannata, Iof\/fe and Nishnianidze (2000) \cite{ACIN00}, Fern\'andez, Negro and Nieto (2004) \cite{FNN04}, Carballo, Fern\'andez, Negro and Nieto (2004) \cite{CFNN04}, Mateo and Negro (2008) \cite{MN08}, Clarkson et al.~\cite{SHC05,FC08}, Marquette \cite{Mar09a,Mar09b}, among others.

The outline of this work is the following. In Sections \ref{section2}--\ref{section5} we present the required background theory, studied earlier by dif\/ferent authors, while Sections \ref{section6},~\ref{section7} contain the main results of this paper. Thus, in the next Section we present a short overview of SUSY QM. The polynomial deformations of the Heisenberg--Weyl algebra will be studied at Section~\ref{section3}. In Section~\ref{section4} we will address the second-order PHA, the determination of the general systems having third-order dif\/ferential ladder operators and their connection with Painlev\'e~IV equation, while in Section~\ref{section5} the SUSY partner potentials of the harmonic oscillator will be analyzed. In Section~\ref{section6} we will formulate a theorem with the requirements that these SUSY partners have to obey in order to generate solutions of $P_{\rm IV}$, and in Section~\ref{section7} we shall explore three relevant solution hierarchies. Our conclusions will be presented at Section~\ref{section8}.

\section{Supersymmetric quantum mechanics}\label{section2}

Let us consider the following standard intertwining relations \cite{AIS93,FFG05,AICD95,BS97,MRO04,Fer10}
\begin{gather*}
 H_i A_i^{+}  = A_i^{+} H_{i-1}, \qquad H_{i-1}A_i^{-}=A_i^{-}H_i,\\
 A_i^{\pm}  = \frac{1}{\sqrt{2}}\left[\mp \frac{d}{d x} + \alpha_i(x,\epsilon_i)\right], \qquad i = 1,\dots,k,
\end{gather*}
where
\begin{gather*}
H_i = -\frac12 \frac{d^2}{d x^2} + V_i(x), \qquad i=0,\dots,k.
\end{gather*}
Hence, the following equations have to be satisf\/ied
\begin{gather}
 \alpha_i'(x,\epsilon_i) + \alpha_i^2(x,\epsilon_i) = 2[V_{i-1}(x) - \epsilon_i],  \label{rei} \\
 V_{i}(x) = V_{i-1}(x) - \alpha_i'(x,\epsilon_i). \label{npi}
\end{gather}

The previous expressions imply that, once it is found a solution $\alpha_i(x,\epsilon_i)$ of the Riccati equation (\ref{rei}) associated to the potential $V_{i-1}(x)$ and the factorization energy $\epsilon_i$, the new potential $V_{i}(x)$ is completely determined by equation~\eqref{npi}. The key point of this procedure is to realize that the Riccati solution $\alpha_i(x,\epsilon_i)$ of the $i$-th equation can be algebraically determined of two Riccati solutions $\alpha_{i-1}(x,\epsilon_{i-1})$, $\alpha_{i-1}(x,\epsilon_{i})$ of the  $(i-1)$-th equation in the way:
\begin{gather*}
\alpha_{i}(x,\epsilon_{i}) = - \alpha_{i-1}(x,\epsilon_{i-1}) - \frac{2(\epsilon_{i-1}- \epsilon_{i})}
{\alpha_{i-1}(x,\epsilon_{i-1}) - \alpha_{i-1}(x,\epsilon_{i})}.
\end{gather*}
By iterating down this equation, it turns out that $\alpha_{i}(x,\epsilon_{i})$ is determined either by $i$ solutions of the initial Riccati equation
\begin{gather*}
\alpha_1'(x,\epsilon_j) + \alpha_1^2(x,\epsilon_j) = 2 [V_0(x) - \epsilon_j], \qquad j=1,\dots,i,
\end{gather*}
or by $i$ solutions $u_j$ of the corresponding Schr\"odinger equation
\begin{gather}
H_0 u_j = - \frac12 u_j'' + V_0(x)u_j = \epsilon_j u_j, \qquad j=1,\dots,i, \label{usch}
\end{gather}
where $\alpha_1(x,\epsilon_j) = u_j'/u_j$.

Now, the $k$-th order supersymmetric quantum mechanics realizes the standard SUSY algebra with two generators
\begin{gather*}
[{\sf Q}_i, {\sf H}_{\rm ss}]=0, \qquad \{ {\sf Q}_i,{\sf Q}_j\} = \delta_{ij} {\sf H}_{\rm ss}, \qquad i,j=1,2,
\end{gather*}
in the way
\begin{gather}
  {\sf Q}_1 =\frac{{\sf Q}^{+} + {\sf Q}^{-}}{\sqrt{2}}, \qquad  {\sf Q}_2 =
\frac{{\sf Q}^{+} - {\sf Q}^{-}}{i\sqrt{2}}, \qquad
  {\sf Q}^{+} = \left(
\begin{matrix}
0 & B_k^{+} \\ 0 & 0
\end{matrix}
\right), \qquad
{\sf Q}^{-} =  \left(
\begin{matrix}
0 & 0 \cr B_k^{-} & 0
\end{matrix}
	\right),	\nonumber\\
   {\sf H}_{\rm ss} = \{Q^{-},Q^{+}\}=
\left(
\begin{matrix}
B_k^{+} B_k^{-}  & 0 \cr 0 & B_k^{-}B_k^{+}
\end{matrix}
\right) = ({\sf
H}_{\rm d}-\epsilon_1) \cdots ({\sf H}_{\rm d}-\epsilon_k), \label{hssfactorized}
\end{gather}
where
\begin{gather*}
{\sf H}_{\rm d} - \epsilon_i = \left(
\begin{matrix}
H_k - \epsilon_i & 0 \cr 0 & H_0 - \epsilon_i
\end{matrix}
\right), \qquad i=1,\dots,k.
\end{gather*}
Notice that, in this formalism, $H_0$ and $H_k$ are the intertwined initial and f\/inal Schr\"odinger Hamiltonians respectively, which means that
\begin{gather}
H_k B_{k}^{+} = B_{k}^{+} H_0, \qquad H_0 B_k^{-}= B_k^{-}H_k, \label{intertwiningk}
\end{gather}
with
\begin{gather*}
B_{k}^{+} = A_k^{+} \dots A_1^{+}, \qquad B_{k}^{-}=A_1^{-}\dots A_k^{-},
\end{gather*}
being $k$-th order dif\/ferential intertwining operators. The initial and f\/inal potentials $V_0$, $V_k$ are related by
\begin{gather*}
V_k(x) = V_0(x) - \sum_{i=1}^k\alpha_i' (x,\epsilon_i) = V_0(x) -
\{\ln[W(u_1,\dots,u_k)]\}'',
\end{gather*}
where $W(u_1,\dots,u_k)$ is the Wronskian of the $k$ Schr\"odinger seed solutions $u_j$, which satisfy equation~\eqref{usch} and are chosen to implement the $k$-th order SUSY transformation.

\section{Polynomial Heisenberg algebras: algebraic properties}\label{section3}

The PHA of $m$-th order\footnote{In this work we will use the terminology of \cite{CFNN04}, although in \cite{AS07} these systems are called of $(m+1)$-th order.} are deformations of the Heisenberg--Weyl algebra \cite{AIS93,VS93,Spi95,FH99}:
\begin{gather*}
[H,L^\pm] 		 = \pm L^\pm , \qquad
[L^-,L^+] 			  \equiv Q_{m+1}(H+1) - Q_{m+1}(H) = P_m(H) , \\
Q_{m+1}(H) 	 = L^+ L^- = \prod\limits_{i=1}^{m+1} \left(H - {\cal E}_i\right) ,
\end{gather*}
where the Hamiltonian describing those systems has the standard Schr\"odinger form
\begin{gather*}
H = -\frac12 \frac{d^2}{d x^2} + V(x) , 
\end{gather*}
$Q_{m+1}(x)$ is a $(m+1)$-th order polynomial in $x$ which implies that $P_m(x)$ is a polynomial of order $m$ in $x$, and in this work we assume that $L^{-}=(L^{+})^{\dagger}$.

\begin{figure}[t]
\centering
\includegraphics[scale=0.2]{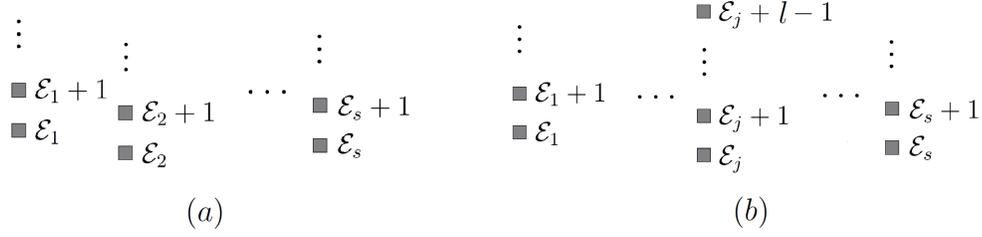}

\caption{Possible spectra for systems characterized by $m$-th order PHA, $s\leq m+1$. In case (\textit{a}), the spectrum consists of $s$ inf\/inite physical ladders, while in case (\textit{b}) the $j$-th ladder has f\/inite length since equation~\eqref{finladcon} is satisf\/ied.}\label{Fig1}
\end{figure}

The corresponding spectrum of $H$, ${\rm Sp}(H)$, is determined by the $s$ physical eigenstates of $H$ belonging as well to the kernel of $L^-$:
\begin{gather*}
L^-\psi_{{\cal E}_i} = 0, \qquad H\psi_{{\cal E}_i} = {\cal E}_i \psi_{{\cal E}_i}, \qquad i= 1,\dots,s,
\end{gather*}
where $s$ is an integer taking a value between $1$ and $m+1$. $s$ independent ladders can be constructed departing from these physical extremal states by acting repeatedly the operator $L^+$ on each $\psi_{{\cal E}_i}$. In general, these ladders are of inf\/inite length. An illustration of this  situation is shown in Fig.~\ref{Fig1}\textit{a}.

On the other hand, for a certain extremal state $\psi_{{\cal E}_j}$ it could happen that
\begin{gather}
\left( L^+\right)^{l-1}\psi_{{\cal E}_j} \neq 0, \qquad \left(L^+\right)^{l}\psi_{{\cal E}_j} = 0, \label{finladcon}
\end{gather}
for some integer $l$. In this case it turns out that one of the roots dif\/ferent from the physical ones $\{{\cal E}_1, \dots, {\cal E}_s\}$ acquires the form
\begin{gather*}
{\cal E}_i = {\cal E}_j + l, \qquad i\in\{s+1, \dots, m + 1\}.
\end{gather*}
Thus, the $j$-th ladder, which departs from ${\cal E}_j$, will end at ${\cal E}_j + l - 1$,
 i.e., it has f\/inite length. An illustration of this situation is shown in Fig.~\ref{Fig1}\textit{b}.

\section{Second-order polynomial Heisenberg algebras}\label{section4}

By making $m=2$ we get now the second-order PHA for which \cite{AIS93,VS93,Spi95,FH99,Spi92}:
\begin{gather}
Q_{3}(H)   = \left(H - {\cal E}_1\right) \left(H - {\cal E}_2\right)\left(H - {\cal E}_3\right), \label{q3}\\
P_2(H)   = 3H^2 + [3-2({\cal E}_1 + {\cal E}_2 + {\cal E}_3)]H + 1 -({\cal E}_1 + {\cal E}_2 + {\cal E}_3) + {\cal E}_1{\cal E}_2 + {\cal E}_1{\cal E}_3 + {\cal E}_2{\cal E}_3 .\nonumber
\end{gather}
From the purely algebraic point of view, the systems ruled by second-order PHA could have up to three independent physical ladders, each one starting from ${\cal E}_1$, ${\cal E}_2$ and ${\cal E}_3$.

Now, let us take a look at the dif\/ferential representation of the second-order PHA. Suppose that $L^+$ is a third-order dif\/ferential ladder operator, chosen by simplicity in the following way:
\begin{gather*}
L^+    = L_a^+ L_b^+ , \qquad
L_a^+  = \frac{1}{\sqrt{2}}\left[-\frac{d}{d x} + f(x) \right], \qquad
L_b^+  = \frac12\left[ \frac{d^2}{d x^2} + g(x)\frac{d}{d x} +
h(x)\right].
\end{gather*}
These operators satisfy the following intertwining relationships:
\begin{gather*}
HL_a^+ = L_a^+ (H_{\rm a} + 1), \qquad H_{\rm a} L_b^+ = L_b^+ H \quad \Rightarrow \quad [H,L^+] = L^+,
\end{gather*}
where $H_{\rm a}$ is an intermediate auxiliary Schr\"odinger Hamiltonian. By using the standard equations for the f\/irst and second-order SUSY QM, it turns out that
\begin{gather*}
  -f' + f^2 = 2(V - {\cal E}_1), \\
  V_{\rm a} = V + f' -1 = V + g',  \\
  \frac{g''}{2g} - \left(\frac{g'}{2g}\right)^2 - g' + \frac{g^2}4 + \frac{({\cal E}_2 - {\cal E}_3)^2}{g^2} + {\cal E}_2 + {\cal E}_3 -2 = 2V,\\
  h = - \frac{g'}{2} + \frac{g^2}{2} - 2V + {\cal E}_2 + {\cal E}_3 - 2.
\end{gather*}
By  decoupling this system it is obtained:
\begin{gather}
  f = x + g, \label{fdependg} \\
  h = - x^2 + \frac{g'}{2} - \frac{g^2}{2} - 2xg + a, \\
  V = \frac{x^2}2 - \frac{g'}2 + \frac{g^2}2 + x g + {\cal E}_1 -
\frac12 , \label{Vpivs}
\end{gather}
where
\begin{gather*}
g'' = \frac{g'^2}{2g} + \frac{3}{2} g^3 + 4xg^2 + 2\left(x^2 - a \right) g + \frac{b}{g}.
\end{gather*}
Notice that this is the Painlev\'e IV equation ($P_{\rm IV}$) with parameters $a ={\cal E}_2 + {\cal E}_3-2{\cal E}_1 -1$, $b = - 2({\cal E}_2 - {\cal E}_3)^2$.

As can be seen, if one solution $g(x)$ of $P_{\rm IV}$ is obtained for certain values of ${\cal E}_1, \ {\cal E}_2, \ {\cal E}_3$, then the potential $V(x)$ as well as the corresponding ladder operators $L^\pm$ are completely determined (see equations~(\ref{fdependg})--(\ref{Vpivs})). Moreover, the three extremal states, some of which could have physical interpretation, are obtained from the following expressions
\begin{gather}
 \psi_{{\cal E}_1}  \propto \exp\left( - \frac{x^2}{2} - \int g dx\right), \label{exes3} \\
\psi_{{\cal E}_2}  \propto \left( \frac{g'}{2g} - \frac{g}{2} - \frac{\Delta}g - x\right)
\exp\left[\int\left( \frac{g'}{2g} + \frac{g}{2} - \frac{\Delta}g \right) dx \right],\nonumber\\ 
\psi_{{\cal E}_3}  \propto \left( \frac{g'}{2g} - \frac{g}{2} + \frac{\Delta}g - x\right)
\exp\left[\int\left( \frac{g'}{2g} + \frac{g}{2} + \frac{\Delta}g \right) dx \right], \nonumber 
\end{gather}
where $\Delta = {\cal E}_2 - {\cal E}_3$. The corresponding physical ladders of our system are obtained departing from the extremal states with physical meaning. In this way we can determine the spectrum of the Hamiltonian~$H$.

On the other hand, if we have identif\/ied a system ruled by third-order dif\/ferential ladder operators, it is possible to design a mechanism for obtaining solutions of the Painlev\'e IV equation. The key point of this procedure is to identify the extremal states of our system; then, from the expression for the extremal state of equation~\eqref{exes3} it is straightforward to see that
\begin{gather*}
g(x) = - x - \{\ln[\psi_{{\cal E}_1}(x)]\}'.
\end{gather*}
Notice that, by making cyclic permutations of the indices of the initially assigned extremal states $\psi_{{\cal E}_1}$, $\psi_{{\cal E}_2}$, $\psi_{{\cal E}_3}$, we obtain three solutions of $P_{\rm IV}$ with dif\/ferent parameters $a$, $b$.

\section[$k$-th order SUSY partners of the harmonic oscillator]{$\boldsymbol{k}$-th order SUSY partners of the harmonic oscillator}\label{section5}

To apply the $k$-th order SUSY transformations to the harmonic oscillator we need the general solution $u(x,\epsilon)$ of the Schr\"odinger equation for $V_0(x) = x^2/2$ and an arbitrary factorization energy~$\epsilon$. A straightforward calculation~\cite{JR98} leads to:
\begin{gather}
u(x,\epsilon )   = e^{-x^2/2}\left[ {}_1F_1\left(\frac{1-2\epsilon}{4},\frac12;x^2\right) + 2x\nu\frac{\Gamma(\frac{3 - 2\epsilon}{4})}{\Gamma(\frac{1-2\epsilon}{4})}\, {}_1F_1\left(\frac{3-2\epsilon}{4},\frac32;x^2\right)\right] \nonumber \\
\hphantom{u(x,\epsilon )}{}  = e^{x^2/2}\left[ {}_1F_1\left(\frac{1+2\epsilon}{4},\frac12;-x^2\right) + 2x\nu\frac{\Gamma(\frac{3 - 2\epsilon}{4})}{\Gamma(\frac{1-2\epsilon}{4})}\, {}_1F_1\left(\frac{3+2\epsilon}{4},\frac32;-x^2\right)\right] , \label{hyper}
\end{gather}
where $_1F_1$ is the conf\/luent hypergeometric (Kummer) function. Note that, for $\epsilon < 1/2$, this solution will be nodeless for $\vert\nu\vert < 1$ while it will have one node for $\vert\nu\vert > 1$.

Let us perform now a non-singular $k$-th order SUSY transformation which creates precisely~$k$ new
levels, additional to the standard ones $E_n = n + 1/2$, $n=0,1,2,\dots$ of $H_0$, in the way
\begin{gather}
\text{Sp}(H_k)  = \left\{ \epsilon_k, \dots,\epsilon_1, \frac12,\frac32,\dots \right\} , \label{spect}
\end{gather}
where $\epsilon_k < \cdots < \epsilon_1 < 1/2$. In order that the Wronskian $W(u_1,\dots,u_k)$ would be nodeless, the parameters $\nu_i$ have to be taken as $\vert\nu_i\vert < 1$ for $i$ odd and $\vert\nu_i\vert > 1$ for $i$ even, $i=1,\dots,k$. The corresponding potential turns out to be
\begin{gather*}
V_k(x) = \frac{x^2}2 - \{\ln[W(u_1,\dots,u_k)]\}''.
\end{gather*}

It is important to note that there is a pair of natural ladder operators $L_k^\pm$ for $H_k$:
\begin{gather*}
L_k^{\pm} = B_k^{+} a^{\pm} B_k^{-},
\end{gather*}
which are dif\/ferential operators of $(2k+1)$-th order such that
\begin{gather*}
[H_k,L_k^\pm]= \pm L_k^\pm ,
\end{gather*}
and $a^{-}$, $a^{+}$ are the standard annihilation and creation operators of the harmonic oscillator.

From the intertwining relations of equation~\eqref{intertwiningk} and the factorizations in equation~\eqref{hssfactorized} it is straightforward to show the following relation:
\begin{gather*}
Q_{2k+1}(H_k) = L_k^+ L_k^- = \left(H_k - \frac12\right) \prod_{i=1}^k \left(H_k - \epsilon_i \right) \left(H_k - \epsilon_i - 1\right). 
\end{gather*}
This means that $\{H_k,  L_k^-,  L_k^+\}$ generate a $(2k)$-th order PHA such that
\begin{gather*}
[L_k^-,L_k^+]=P_{2k}(H_k).
\end{gather*}
Since the roots of $Q_{2k+1}(x)$ are $\{\epsilon_k, \dots, \epsilon_1, 1/2, \epsilon_k + 1, \dots,\epsilon_1 + 1\}$, it turns out that ${\rm Sp}(H_k)$ contains $k$ one-step ladders, starting and ending at $\epsilon_j$, $j=1,\dots,k$, plus an inf\/inite ladder which departs from $1/2$ (compare with equation~\eqref{spect}).

Note that, for $k=1$ the dif\/ferential ladder operators $L_1^{\pm} = B_1^{+} a^{\pm}B_1^{-}$ are of third order, generating thus a second-order PHA. On the other hand, for $k=2$ it turns out that $L_2^{\pm} = B_2^{+} a^{\pm} B_2^{-}$ are of f\/ifth order, and they generate a fourth-order PHA, etc.

\section{Solutions of the Painlev\'e IV equation through SUSY QM}\label{section6}

It was pointed out at the end of Section~\ref{section4} the way in which solutions of $P_{\rm IV}$ can be generated. Let us now employ this proposal for generating several families of such solutions.

\subsection{First-order SUSY QM}\label{section6.1}

\begin{figure}[h]
\centering
\includegraphics[scale=0.37]{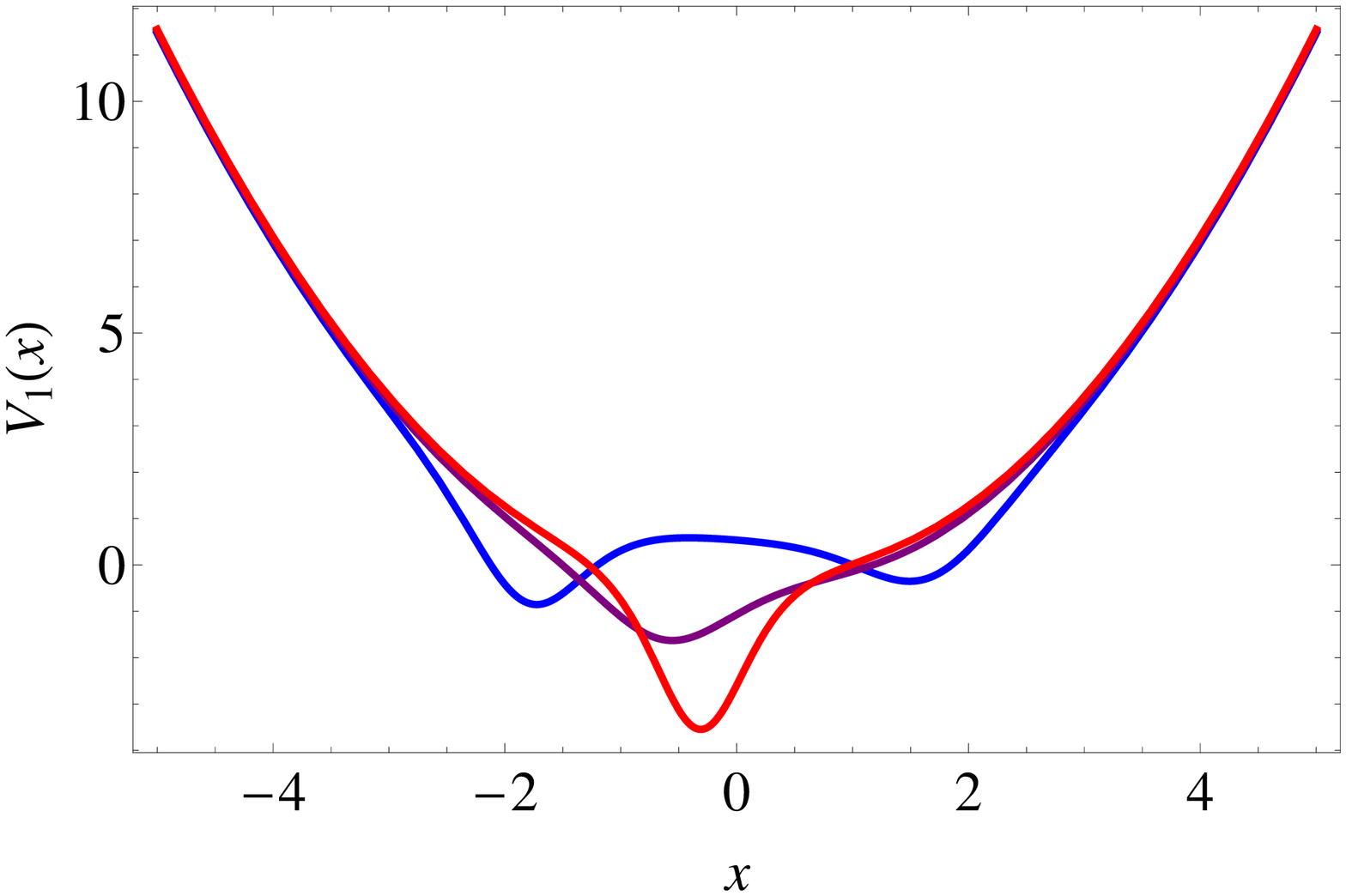} \hskip0.5cm
\includegraphics[scale=0.37]{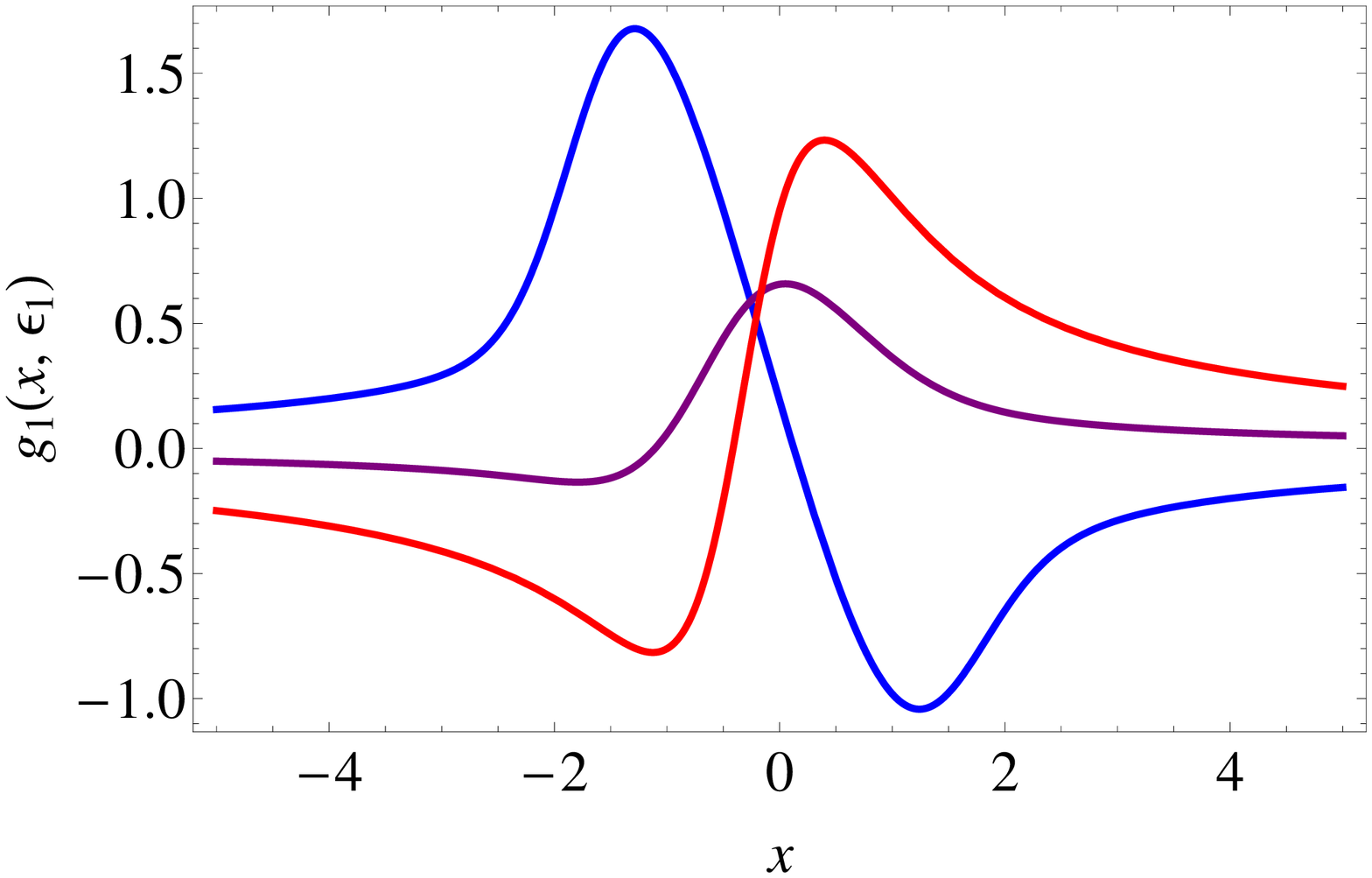}

\caption{First-order SUSY partner potentials $V_1(x)$ (left) of the oscillator and the $P_{\rm IV}$ solutions $g_1(x,\epsilon_1)$ (right) for $\epsilon_1 = \{1/4 \text{(blue)}, -3/4 \text{(purple)}, -7/4 \text{(red)}\}$ and $\nu_1 = 1/2$.}\label{Fig2}
\end{figure}

We saw that, for $k=1$, the ladder operators $L_1^\pm$ are of third order. This means that the f\/irst-order SUSY transformation applied to the oscillator could provide solutions to the $P_{\rm IV}$ equation. To f\/ind them, f\/irst we need to identify the extremal states, which are annihilated by $L_1^-$ and at the same time are eigenstates of $H_1$. From the corresponding spectrum, one realizes that the transformed ground state of $H_0$ and the eigenstate of $H_1$ associated to $\epsilon_1$ are two physical extremal states associated to our system. Since the third root of $Q_3(x)$ is $\epsilon_1 + 1 \not\in{\rm Sp}(H_1)$, then  the corresponding extremal state will be nonphysical, which can be simply constructed from the nonphysical seed solution used to implement the transformation in the way $A_1^+ a^{+} u_1$. Due to this, the three extremal states for our system and their corresponding factorization energies (see equation~\eqref{q3}) become
\begin{gather*}
\psi_{{\cal E}_1} \propto \frac{1}{u_1}, \qquad
\psi_{{\cal E}_2} \propto A_1^+ e^{-x^2/2}, \qquad
\psi_{{\cal E}_3} \propto A_1^+ a^{+} u_1,
\end{gather*}
and
\begin{gather*}
{\cal E}_1 = \epsilon_1, \qquad {\cal E}_2 = \frac{1}{2}, \qquad {\cal E}_3 = \epsilon_1 + 1.
\end{gather*}
The f\/irst-order SUSY partner potential $V_1(x)$ of the oscillator and the corresponding non-singular solution of $P_{\rm IV}$ are
\begin{gather}
V_1(x)  = \frac{x^2}2 - \{\ln [u_1(x)]\}'', \qquad
 g_1(x,\epsilon_1)  =  - x - \{\ln [\psi_{{\cal E}_1}(x)]\}' = - x  + \{\ln[u_1(x)]\}',\label{solg1}
\end{gather}
where we label the $P_{\rm IV}$ solution with an index characterizing the order of the transformation employed and we indicate explicitly the dependence on the factorization energy.  Notice that two additional solutions of the $P_{\rm IV}$ can be obtained by cyclic permutations of the indices $(1,2,3)$. However, they will have singularities at some points and thus we drop them in this approach. An illustration of the f\/irst-order SUSY partner potentials $V_1(x)$ of the oscillator as well as the corresponding solutions $g_1(x,\epsilon_1)$ of $P_{\rm IV}$ for $\epsilon_1 = \{1/4 \text{(blue)}, -3/4 \text{(purple)}, -7/4 \text{(red)}\}$ and $\nu_1 = 1/2$ are shown in Fig.~\ref{Fig2}.

On the other hand, for $k>1$ it is not clear that we can generate solutions of $P_{\rm IV}$. In the next subsection it will be shown that this is possible by imposing certain conditions on the seed solutions used to implement the SUSY transformation.

\subsection[$k$-th order SUSY QM]{$\boldsymbol{k}$-th order SUSY QM}\label{section6.2}

In Section~\ref{section5} we saw that the $k$-th order SUSY partners of the harmonic oscillator are ruled by $(2k)$-th order PHA. It is important to know if there is a way for this algebraic structure to be reduced to a second-order PHA and, if so, which are the requirements. The answer is contained in the following theorem.

\medskip

\noindent {\bf Theorem.} {\it Suppose that the $k$-th order SUSY partner $H_k$ of the harmonic oscillator Hamiltonian $H_0$ is generated by $k$ Schr\"odinger seed solutions $u_j$, $j = 1,\dots,k$ which are connected by the standard annihilation operator in the way:{\samepage
\begin{gather}
u_j = (a^{-})^{j-1} u_1, \qquad \epsilon_j = \epsilon_1 - (j-1), \qquad j=1,\dots,k, \label{restr}
\end{gather}
where $u_1(x)$ is a nodeless Schr\"odinger seed solution given by equation~\eqref{hyper} for $\epsilon_1 < 1/2$ and $\vert \nu_1 \vert < 1$.}

Therefore, the natural ladder operator $L_k^+ = B_k^{+} a^{+} B_k^{-}$ of $H_k$, which is of $(2k+1)$-th order, is factorized in the form
\begin{gather}
L_k^+ = P_{k-1}(H_k) l_k^+,\label{hipo}
\end{gather}
where $P_{k-1}(H_k) = (H_k - \epsilon_1)\cdots(H_k - \epsilon_{k-1})$ is a polynomial of $(k-1)$-th order in $H_k$, $l_k^+$ is a~third-order differential ladder operator such that $[H_k,l_k^+] = l_k^+$ and}
\begin{gather} \label{annumk3}
 l_k^+ l_k^- = (H_k - \epsilon_k)\left(H_k - \frac{1}{2} \right)(H_k - \epsilon_1 - 1).
\end{gather}

\begin{proof}[Proof (by induction)] For $k=1$ the result is obvious since
\begin{gather*}
 L_1^+ = P_0(H_1)l_1^+ , \qquad P_0(H_1) = 1.
\end{gather*}

Let us suppose now that the theorem is valid for a given $k$; then, we are going to show that it is as well valid for $k+1$. From the intertwining technique it is clear that we can go from $H_k$ to $H_{k+1}$ and vice versa through a f\/irst-order SUSY transformation
\begin{gather*}
H_{k+1} A_{k+1}^+ = A_{k+1}^+ H_k, \qquad H_kA_{k+1}^{-}=A_{k+1}^{-}H_k.
\end{gather*}
Moreover, it is straightforward to show that
\begin{gather*}
L_{k+1}^+ = A_{k+1}^+ L_{k}^+ A_{k+1}^- .
\end{gather*}
By using now the induction hypothesis of equation~\eqref{hipo} for the index $k$ it turns out that
\begin{gather}\label{lkm1ell}
L_{k+1}^+ = A_{k+1}^+ P_{k-1}(H_k) l_k^+ A_{k+1}^- = P_{k-1}(H_{k+1})A_{k+1}^+ l_k^+ A_{k+1}^- \equiv P_{k-1}(H_{k+1}) \ell_{k+1}^+,
\end{gather}
where $\ell_{k+1}^+ = A_{k+1}^+ l_k^+ A_{k+1}^-$ is a f\/ifth-order dif\/ferential ladder operator for $H_{k+1}$. It is straightforward to show
\begin{gather*}
\ell_{k+1}^+ \ell_{k+1}^- = (H_{k+1} - \epsilon_k)^2 (H_{k+1} - \epsilon_{k+1})\left(H_{k+1} - \frac{1}{2}\right) (H_{k+1} - \epsilon_1 - 1).
\end{gather*}
Note that the term $(H_{k+1} - \epsilon_{k+1})\left(H_{k+1} - \frac{1}{2}\right)(H_{k+1} - \epsilon_1 - 1)$ in this equation is precisely the result that would be obtained from the product $l_{k+1}^+ l_{k+1}^-$ for the third-order ladder operators of~$H_{k+1}$. Thus, it is concluded that
\begin{gather*}
\ell_{k+1}^+ = q(H_{k+1}) l_{k+1}^+
\end{gather*}
where $q(H_{k+1})$ is a polynomial of $H_{k+1}$. By remembering that $\ell_{k+1}^+$, $l_{k+1}^+$ and $H_{k+1}$ are dif\/ferential operators of 5-th, 3-th, and 2-th order respectively, one can conclude that $q(H_{k+1})$ is lineal in $H_{k+1}$ and therefore
\begin{gather*}
\ell_{k+1}^+ = (H_{k+1} - \epsilon_k) l_{k+1}^+ .
\end{gather*}
By substituting this result in equation~\eqref{lkm1ell} we f\/inally obtain:
\begin{gather*}
L_{k+1}^+ = P_{k-1}(H_{k+1})(H_{k+1} - \epsilon_k) l_{k+1}^+ = P_{k}(H_{k+1})l_{k+1}^+ .  \tag*{\qed}
\end{gather*}
\renewcommand{\qed}{}
\end{proof}

We have determined the restrictions on the Schr\"odinger seed solutions $u_j$ to reduce the order of the natural algebraic structure of the Hamiltonian $H_k$ from $2k$ to $2$. Now suppose we stick to these constraints for generating $H_k$. Since the reduced ladder operator $l_k^{+}$ is of third order, it turns out that we can once again obtain solutions of the Painlev\'e IV equation. To get them, we need to identify the extremal states of our system. Since the roots of the polynomial of equation~\eqref{annumk3} are $\epsilon_k = \epsilon_1 - (k-1),\, 1/2,\, \epsilon_1 + 1$, the spectrum of $H_k$ consists of two physical ladders: a f\/inite one starting from $\epsilon_k$ and ending at $\epsilon_1$; an inf\/inite one departing from $1/2$. Thus, the two physical extremal states correspond to the eigenstate of $H_k$ associated to $\epsilon_k $ and to the mapped eigenstate of $H_0$ with eigenvalue $1/2$. The third extremal state (which corresponds to $\epsilon_1 +1 \not\in{\rm Sp}(H_k)$) is nonphysical, proportional to $B_k^{+} a^{+} u_1$. Thus, the three extremal states are
\begin{gather*}
\psi_{{\cal E}_1} \propto \frac{W(u_1,\dots,u_{k-1})}{W(u_1,\dots,u_k)}, \qquad
\psi_{{\cal E}_2} \propto B_k^+ e^{-x^2/2}, \qquad
\psi_{{\cal E}_3} \propto B_k^+ a^{+} u_1,
\end{gather*}
with
\begin{gather*}
{\cal E}_1 = \epsilon_k = \epsilon_1 - (k - 1),  \qquad {\cal E}_2 = \frac{1}{2}, \qquad {\cal E}_3 = \epsilon_1 + 1.
\end{gather*}
The $k$-th order SUSY partner of the oscillator potential and the corresponding non-singular solution of the $P_{\rm IV}$ equation become
\begin{gather} 
V_k(x)  = \frac{x^2}2 - \{\ln [W(u_1,\dots,u_k)]\}'' ,\nonumber \\
g_k(x,\epsilon_1)  = - x - \{\ln[\psi_{{\cal E}_1}(x)]\}' =  - x - \left\{\ln \left[\frac{W(u_1,\dots,u_{k-1})}{W(u_1,\dots,u_{k})}\right]\right\}', \qquad k\geq 2.\label{solg}
\end{gather}

Let us remind that the $k$ Schr\"odinger seed solutions in the previous expressions are not longer arbitrary; they have to obey the restrictions imposed by our theorem (see equation~\eqref{restr}).

\begin{figure}[t]
\centering
\includegraphics[scale=0.37]{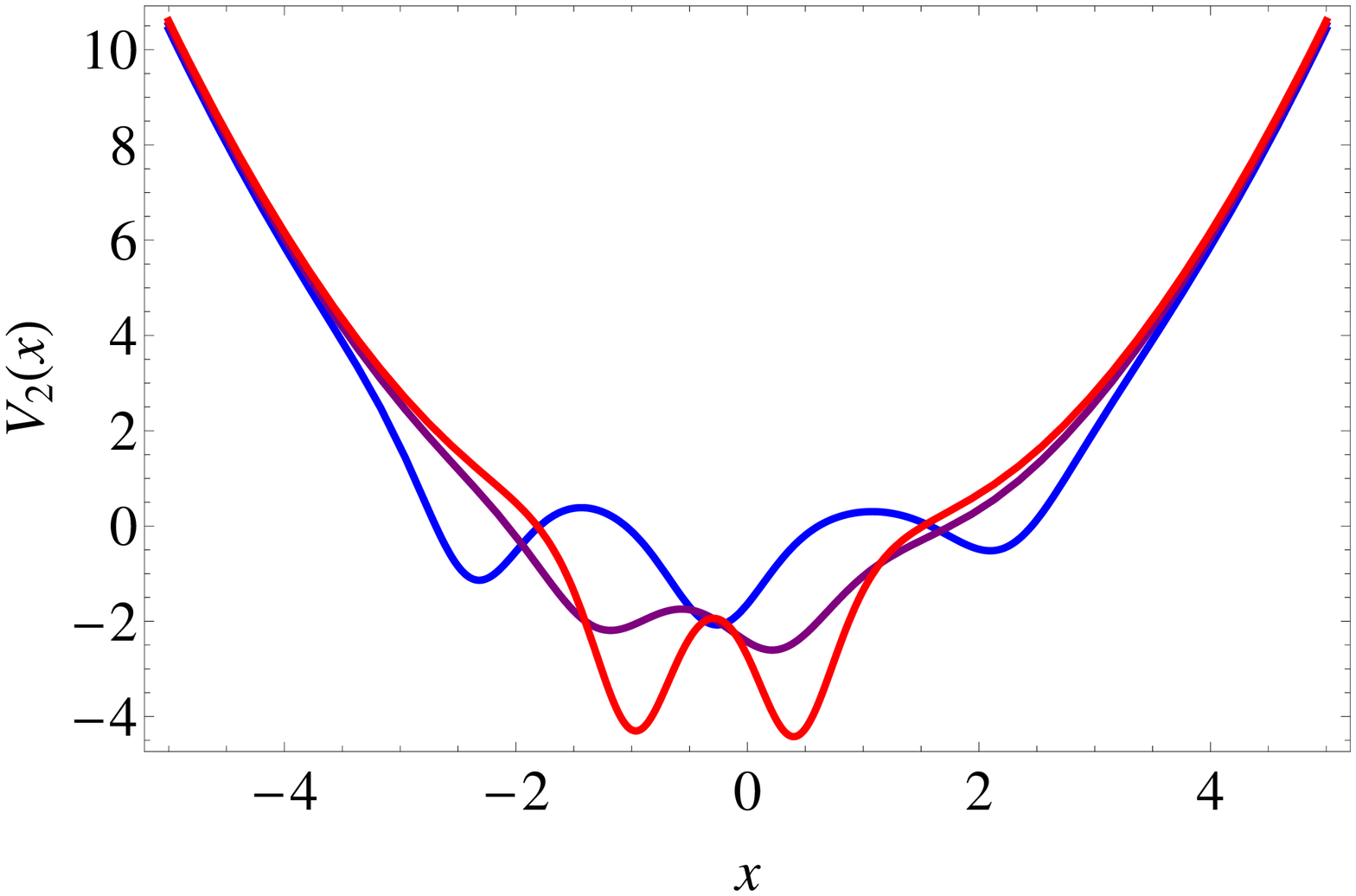} \hskip0.5cm
\includegraphics[scale=0.37]{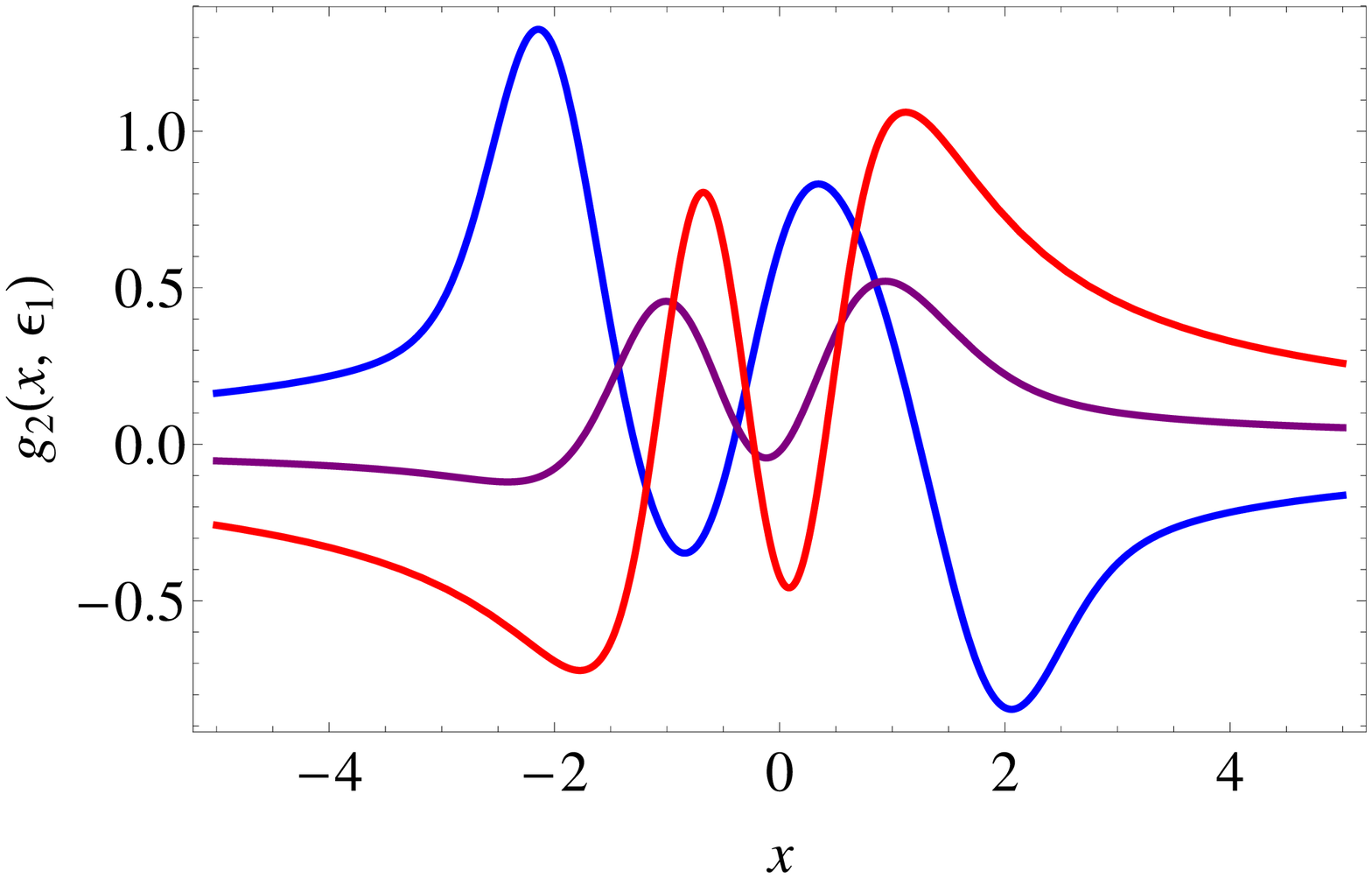}

\caption{Second-order SUSY partner potentials $V_2(x)$ (left) of the oscillator and the correspon\-ding~$P_{\rm IV}$ solutions $g_2(x,\epsilon_1)$ (right) for $\epsilon_1 = \{1/4 \text{(blue)}, -3/4 \text{(purple)}, -7/4 \text{(red)}\}$ and $\nu_1 = 1/2$.}\label{Fig3}
\end{figure}

\begin{figure}[t]
\centering
\includegraphics[scale=0.37]{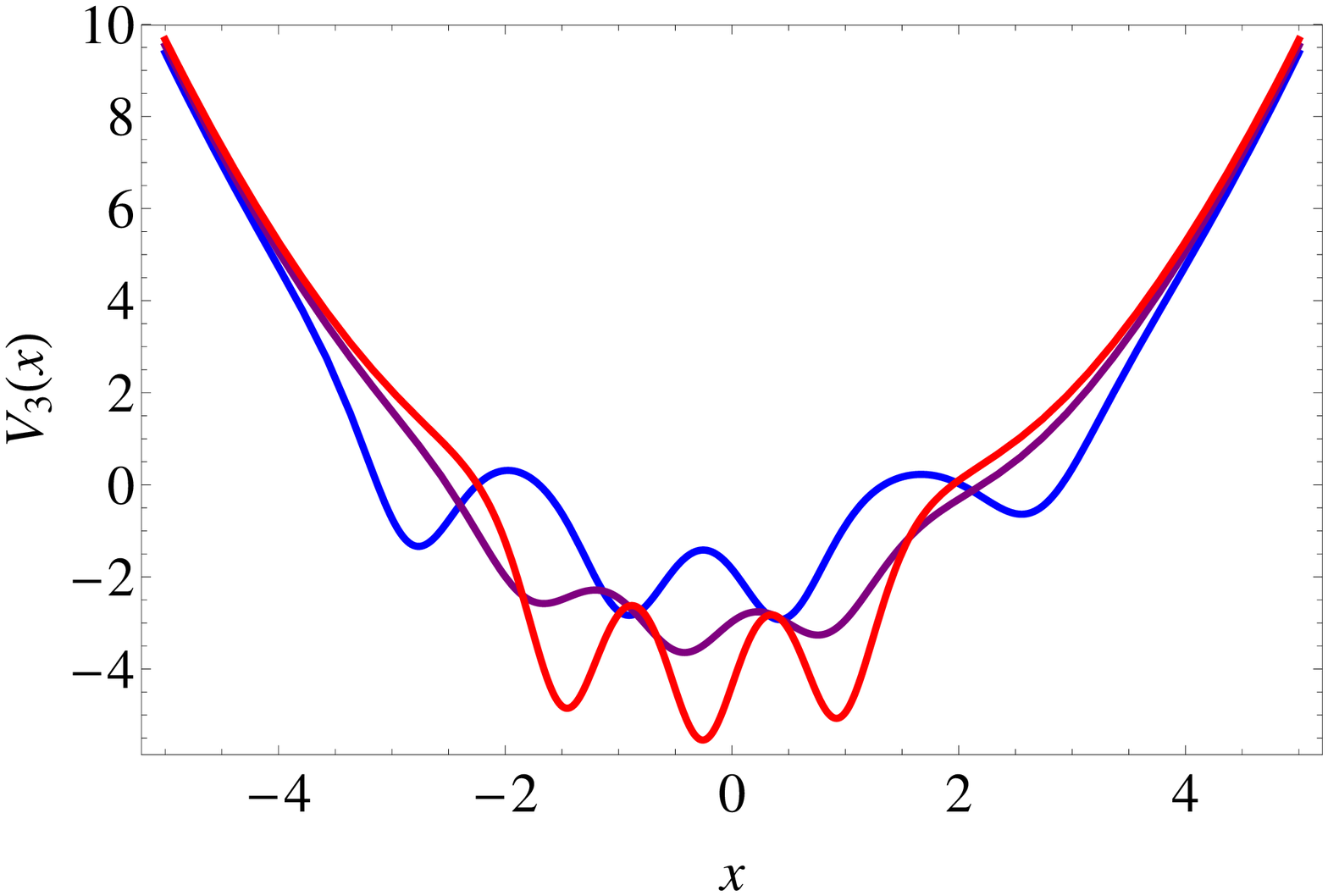} \hskip0.5cm
\includegraphics[scale=0.37]{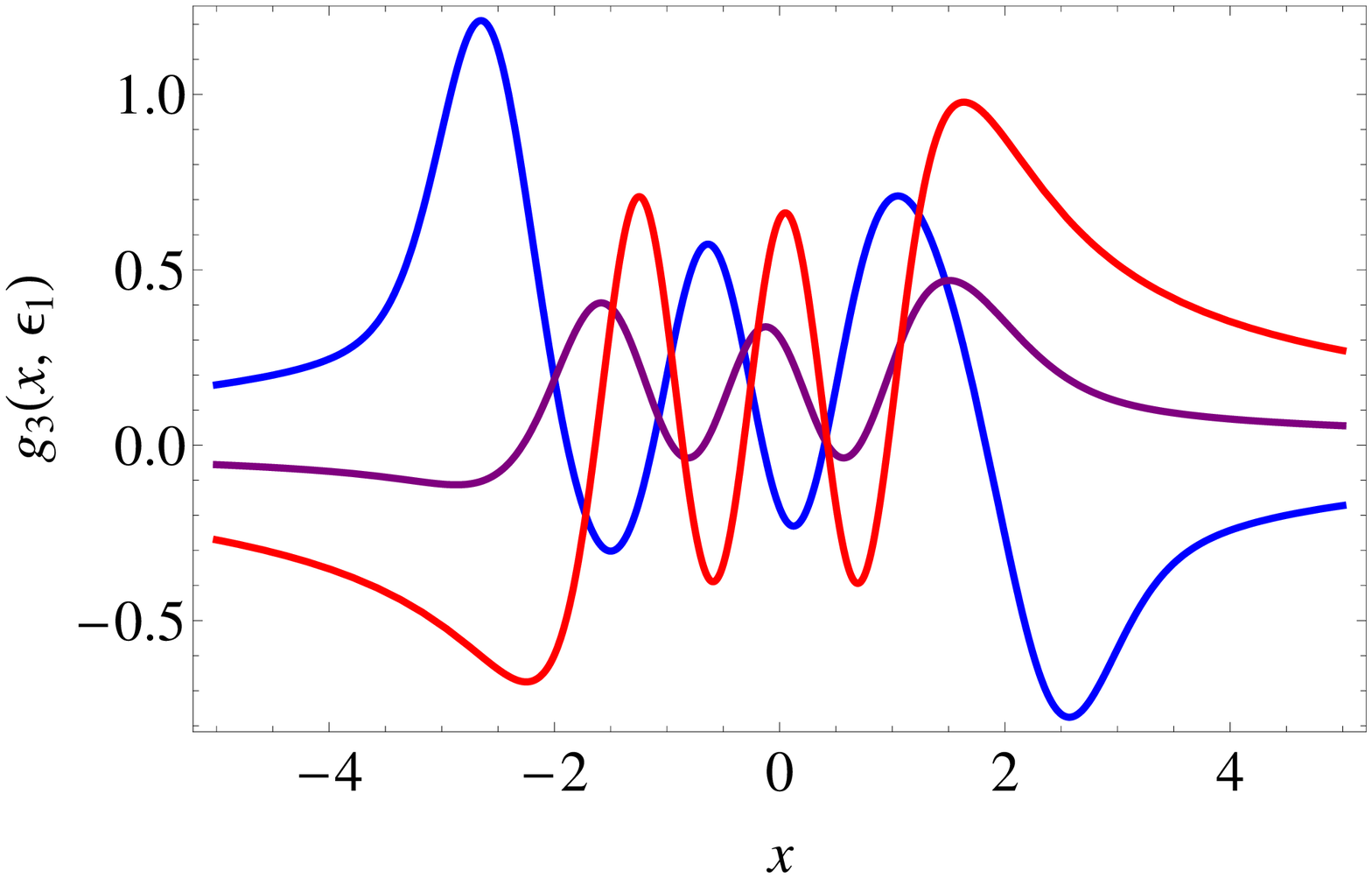}

\caption{Third-order SUSY partner potentials $V_3(x)$ (left) of the oscillator and the correspon\-ding~$P_{\rm IV}$ solutions $g_3(x,\epsilon_1)$ (right) for $\epsilon_1 = \{1/4 \text{(blue)}$, $-3/4 \text{(purple)}$, $-7/4 \text{(red)}\}$ and $\nu_1 = 1/2$.}\label{Fig4}
\end{figure}

We have illustrated the $k$-th order SUSY partner potentials $V_k(x)$ of the oscillator and the corresponding $P_{\rm IV}$ solutions $g_k(x,\epsilon_1)$ in Fig.~\ref{Fig3} ($k=2$) and Fig.~\ref{Fig4} ($k=3$) for $\epsilon_1 = \{1/4 \text{(blue)}$, $-3/4 \text{(purple)}, -7/4 \text{(red)}\}$ and $\nu_1 = 1/2$.

\section{Solution hierarchies. Explicit formulas}\label{section7}

The solutions $g_k(x,\epsilon_1)$ of the Painlev\'e IV equation can be classif\/ied according to the explicit functions of which they depend on~\cite{BCH95}. Our general formulas, given by equations~(\ref{solg1}), (\ref{solg}), in general are expressed in terms of the conf\/luent hypergeometric function $_1F_1$, although for specif\/ic values of the parameter~$\epsilon_1$ they can be simplif\/ied to the error function~$\text{erf}(x)$. Moreover, for particular parameters~$\epsilon_1$ and~$\nu_1$, they simplify further to rational solutions.

Let us remark that, in this paper, we are interested in non-singular SUSY partner potentials and the corresponding non-singular solutions of $P_{\rm IV}$. To accomplish this, we restrict the parameters to $\epsilon_1 <1/2$ and $\vert \nu_1 \vert < 1$.

\subsection{Conf\/luent hypergeometric function hierarchy}\label{section7.1}

In general, the solutions of $P_{\rm IV}$ given in equations~(\ref{solg1}), (\ref{solg}) are expressed in terms of two conf\/luent hypergeometric functions. Next we write down the explicit formula for $g_1(x,\epsilon_1)$ in terms of the parameters $\epsilon_1$, $\nu_1$ (with $\epsilon_1 < 1/2$ and $\vert \nu_1 \vert < 1$ to avoid singularities):
\begin{gather}
g_1(x,\epsilon_1)=   \frac{2\nu_1\Gamma\left(\frac{3-2\epsilon_1}{4}\right) \left[(3-6x^2)\,{}_1F_1\left(\frac{3-2\epsilon_1}{4},\frac{3}{2};x^2\right)+x^2(3-2\epsilon_1)\,{}_1F_1\left(\frac{7-2\epsilon_1}{4},\frac{5}{2};x^2\right) \right]}
{3 \Gamma\left(\frac{1-2\epsilon_1}{4}\right)\,{}_1F_1\left(\frac{1-2\epsilon_1}{4},\frac{1}{2};x^2\right) + 6\nu_1 x \Gamma\left(\frac{3-2\epsilon_1}{4}\right)\,{}_1F_1\left(\frac{3-2\epsilon_1}{4},\frac{3}{2};x^2\right) }\nonumber\\
\hphantom{g_1(x,\epsilon_1)=}{}
+\frac{3x\Gamma\left(\frac{1-2\epsilon_1}{4}\right)\left[ -2\,{}_1F_1\left(\frac{1-2\epsilon_1}{4},\frac{1}{2};x^2\right)+(1-2\epsilon_1)\,{}_1F_1\left(\frac{5-2\epsilon_1}{4},\frac{3}{2};x^2\right)	 \right]}
{3 \Gamma\left(\frac{1-2\epsilon_1}{4}\right)\,{}_1F_1\left(\frac{1-2\epsilon_1}{4},\frac{1}{2};x^2\right) + 6\nu_1 x \Gamma\left(\frac{3-2\epsilon_1}{4}\right)\,{}_1F_1\left(\frac{3-2\epsilon_1}{4},\frac{3}{2};x^2\right) }. \label{g1}
\end{gather}

Notice that the solutions $g_1(x,\epsilon_1)$ plotted in Fig.~\ref{Fig2} for specif\/ic values of $\epsilon_1$, $\nu_1$ correspond to particular cases of this hierarchy. The explicit analytic formulas for higher-order solutions $g_k(x,\epsilon_1)$ can be obtained through formula~\eqref{solg}, and they have a similar form as in equation~\eqref{g1}.

\subsection{Error function hierarchy}\label{section7.2}
It would be interesting to analyze the possibility of reducing the explicit form of the $P_{\rm IV}$ solution to the error function. To do that, let us f\/ix the factorization energy in such a way that any of the two hypergeometric series of equation~\eqref{hyper} reduces to that function. This can be achieved for $\epsilon_1$ taking a value in the set
\begin{gather}
\left\{-\frac12,-\frac32,-\frac52,\dots ,-\frac{(2m+1)}{2},\dots \right\}.\label{ener}
\end{gather}
If we def\/ine the auxiliary function $\varphi_{\nu_1}(x)\equiv \sqrt{\pi}e^{x^2}[1+\nu_1\, \text{erf}(x)]$ to simplify the formulas, we can get simple expressions for $g_k(x,\epsilon_1)$ with some specif\/ic parameters $k$ and $\epsilon_1$:
\begin{gather}
g_1(x,-1/2) =\frac{2\nu_1}{\varphi_{\nu_1}(x)}, \label{erf1}\\
g_1(x,-3/2) =\frac{\varphi_{\nu_1}(x)}{1+x\varphi_{\nu_1}(x)},\nonumber\\
g_1(x,-5/2) =\frac{4[\nu_1 + \varphi_{\nu_1}(x)]}{2\nu_1 x +(1+2x^2)\varphi_{\nu_1}(x)},\label{erf2}\\
g_2(x,-1/2) =\frac{4\nu_1[\nu_1 + 6\varphi_{\nu_1}(x)]}{\varphi_{\nu_1}(x)[\varphi_{\nu_1}^2(x) -2\nu_1 x \varphi_{\nu_1}(x) -2\nu_1^2]}.\nonumber
\end{gather}

\begin{figure}[t]
\centering
\includegraphics[scale=0.37]{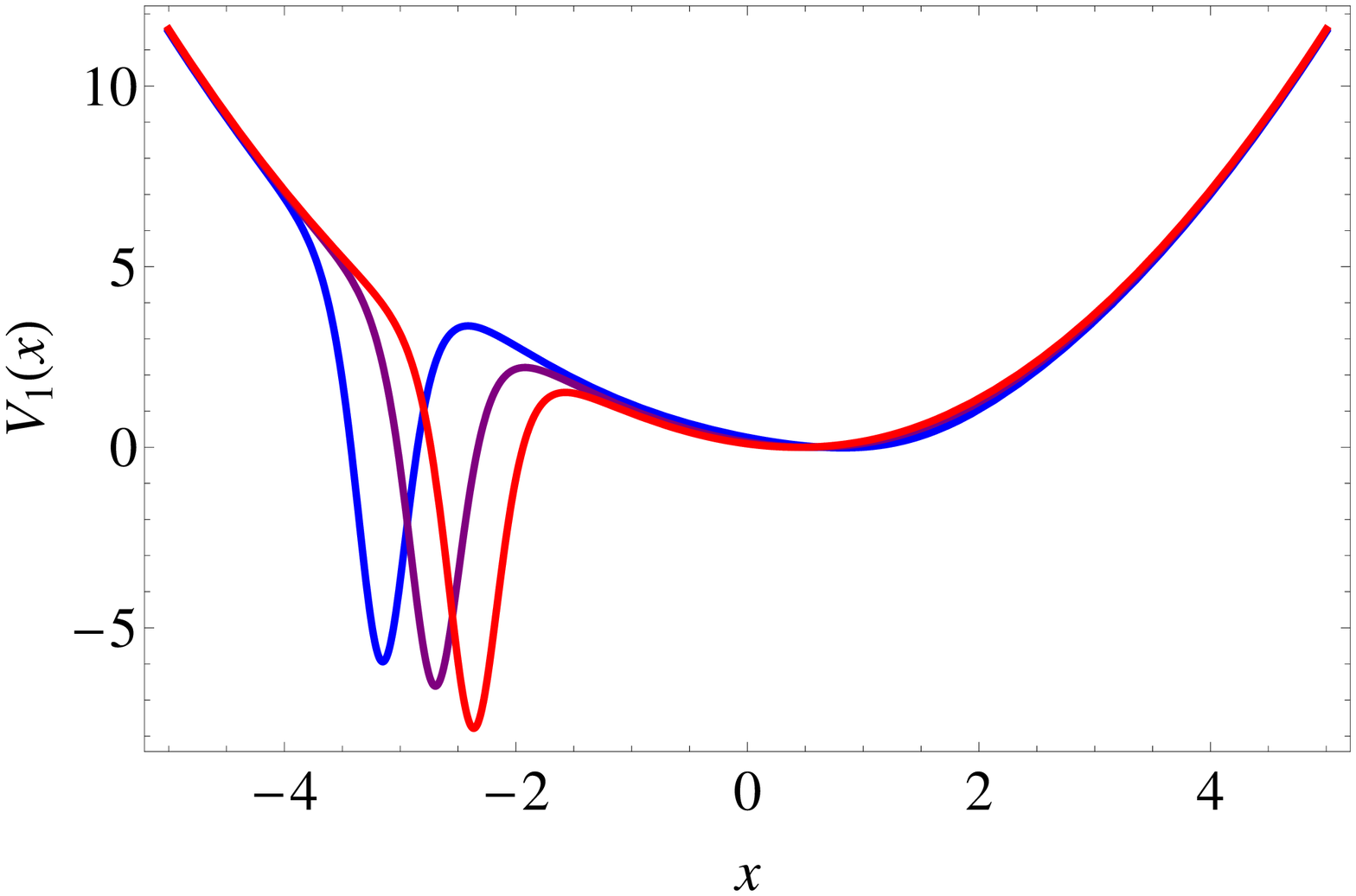} \hskip0.5cm
\includegraphics[scale=0.37]{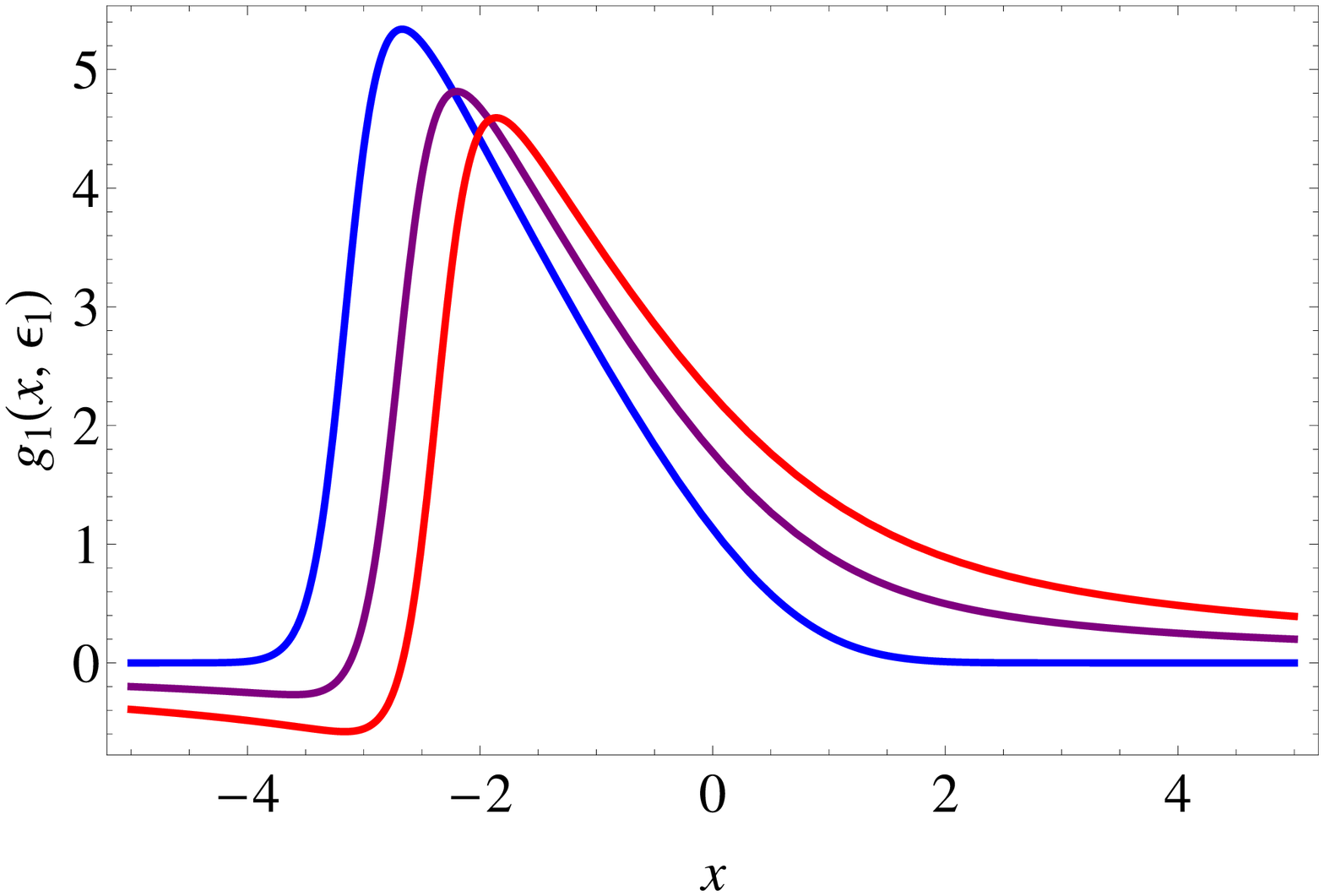}

\caption{First-order SUSY partner potentials $V_1(x)$ (left) of the oscillator and the $P_{\rm IV}$ solutions $g_1(x,\epsilon_1)$ (right) for $\epsilon_1 = \{-1/2 \text{(blue)}, -3/2 \text{(purple)}, -5/2 \text{(red)}\}$ and $\nu_1 = 0.999$, which belong to the error function hierarchy of solutions.}\label{Fig5}
\end{figure}

An illustration of the f\/irst-order SUSY partner potentials $V_1(x)$ of the oscillator and the corresponding $P_{\rm IV}$ solutions $g_1(x,\epsilon_1)$ of equations~(\ref{erf1}), (\ref{erf2}) is given in Fig.~\ref{Fig5}.

\subsection{Rational hierarchy}\label{section7.3}

\begin{figure}[t]
\centering
\includegraphics[scale=0.37]{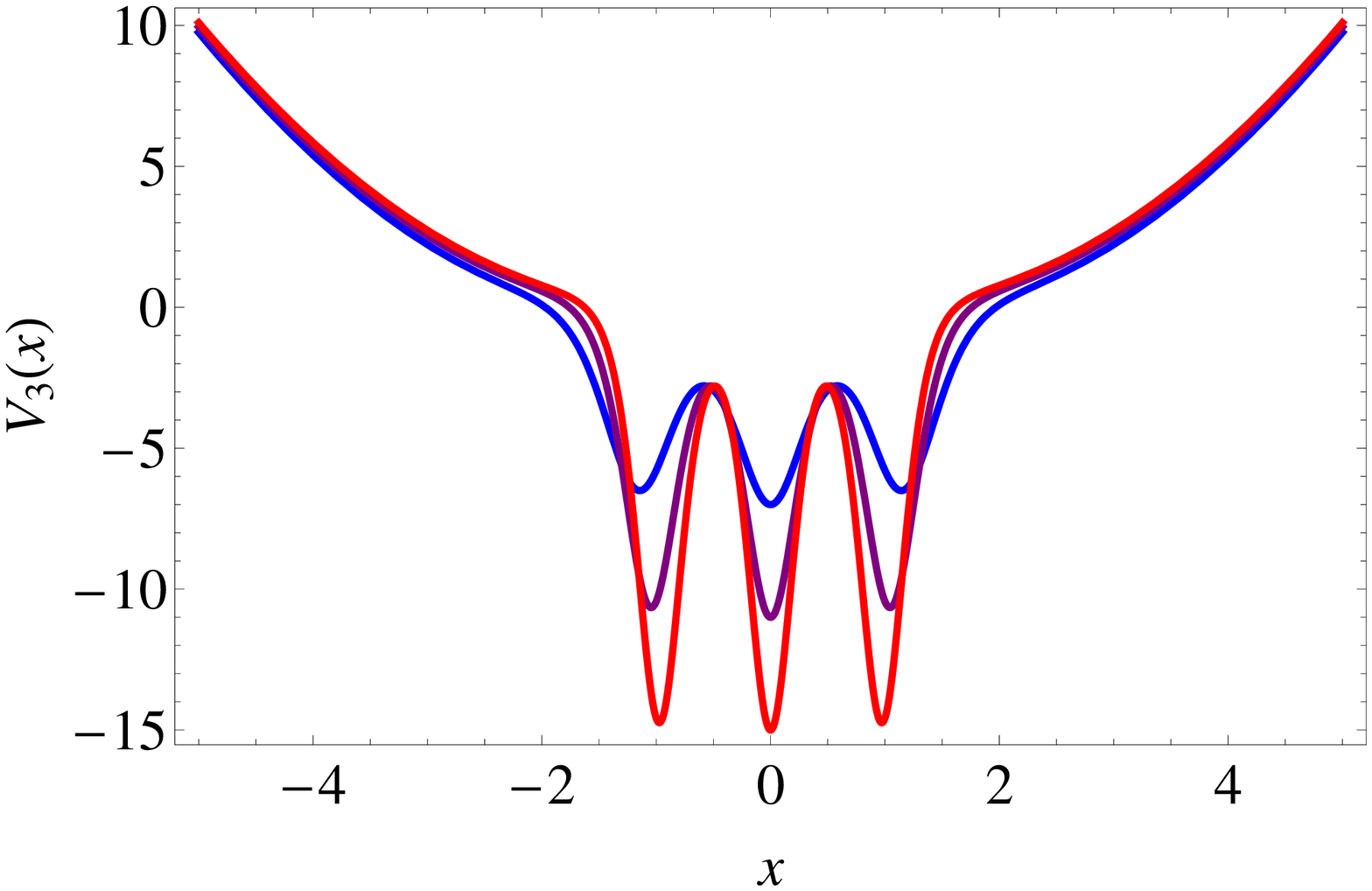} \hskip0.5cm
\includegraphics[scale=0.37]{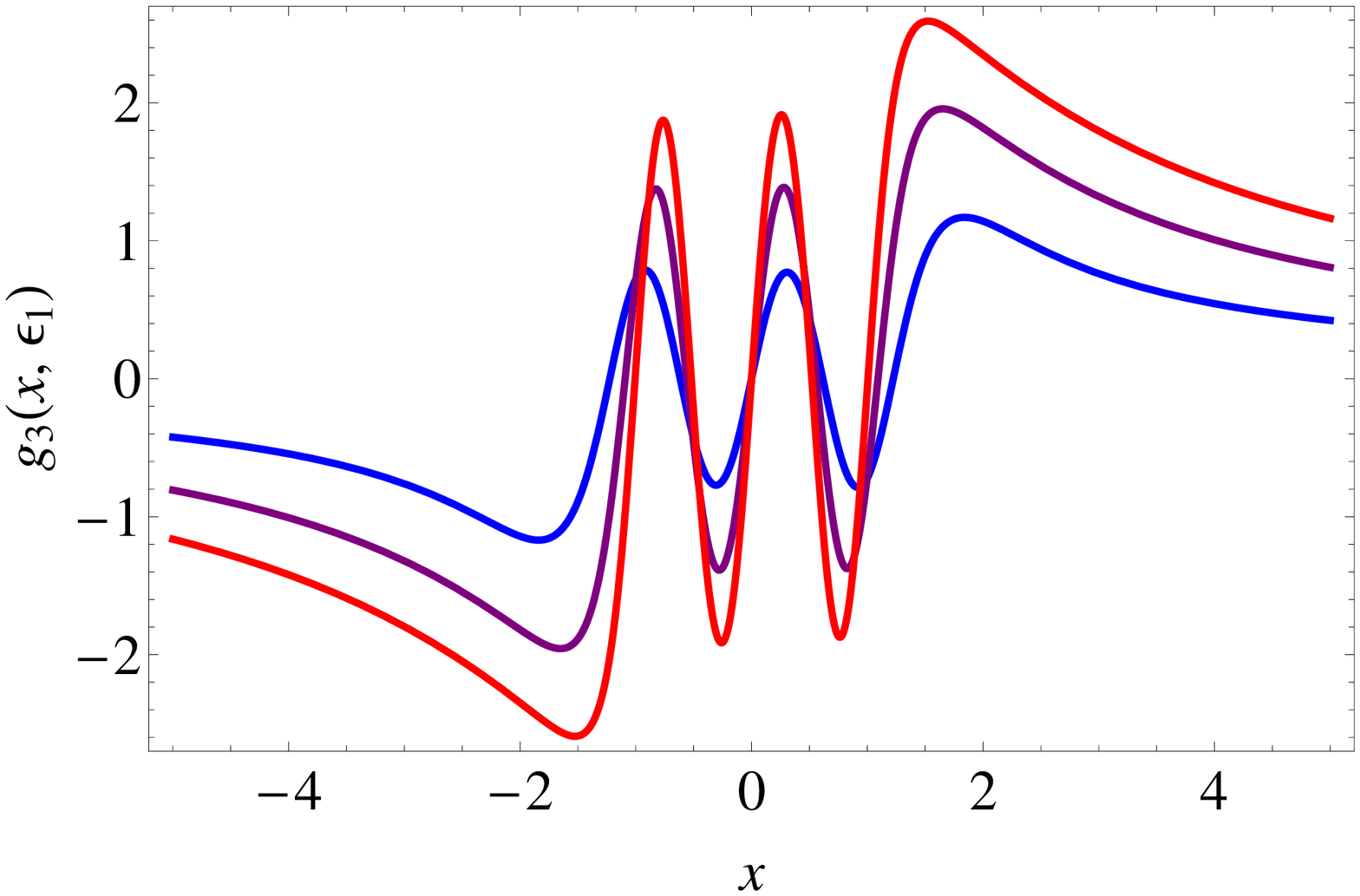}

\caption{Third-order SUSY partner potentials $V_3(x)$ (left) of the oscillator and the $P_{\rm IV}$ solutions $g_3(x,\epsilon_1)$ (right) for $\epsilon_1 = \{-5/2 \text{(blue)}, -9/2 \text{(purple)}, -13/2 \text{(red)}\}$ and $\nu_1 = 0$.}\label{Fig6}
\end{figure}

Our previous formalism allows us to generate solutions of $P_{\rm IV}$ involving in general the conf\/luent hypergeometric series, which has an inf\/inite sum of terms. Let us look for the restrictions needed to reduce the explicit form of equation~\eqref{solg} to non-singular rational solutions. To achieve this, once again the factorization energy $\epsilon_1$ has to take a value in the set given by equation~\eqref{ener}, but depending on the $\epsilon_1$ taken, just one of the two hypergeometric functions is reduced to a~polynomial. Thus, we need to choose additionally the parameter $\nu_1=0$ or $\nu_1\rightarrow\infty$ to keep the appropriate hypergeometric function.
However, for the values $-(4m-1)/2$, $m=1,2,\dots$ and $\nu_1\rightarrow \infty$, it turns out that $u_1$ will have always a node at $x=0$, which will produce one singularity for the corresponding $P_{\rm IV}$ solution. In conclusion, the rational non-singular solutions $g_k(x,\epsilon_1)$ of the $P_{\rm IV}$ arise by making in equation~\eqref{hyper} $\nu_1 = 0$ and
\begin{gather*}
\epsilon_1 \in \left\{-\frac12,-\frac52,\dots, -\frac{(4m+1)}2, \dots \right\}.
\end{gather*}
Taking as the point of departure Schr\"odinger solutions with these $\nu_1$ and $\epsilon_1$ and using our previous expressions~\eqref{solg} for a given order of the transformation we get the following explicit expressions for $g_k(x,\epsilon_1)$, some of which are illustrated in Fig.~\ref{Fig6}:
\begin{gather*}
g_1(x,-5/2)=\frac{4 x}{1 + 2 x^2},\\
g_1(x,-9/2)= \frac{8 (3 x + 2 x^3)}{3 + 12 x^2 + 4 x^4},\\
g_1(x,-13/2)= \frac{12 (15 x + 20 x^3 + 4 x^5)}{15 + 90 x^2 + 60 x^4 + 8 x^6},\\
g_2(x,-5/2)= -\frac{4 x}{1 + 2 x^2} + \frac{16 x^3}{3 + 4 x^4},\\
g_2(x,-9/2)= -\frac{8 (3 x + 2 x^3)}{3 + 12 x^2 + 4 x^4} + \frac{32 (15 x^3 + 12 x^5 + 4 x^7)}{45 + 120 x^4 + 64 x^6 + 16 x^8},\\
g_2(x,-13/2)= -\frac{12 (15 x + 20 x^3 + 4 x^5)}{15 + 90 x^2 + 60 x^4 + 8 x^6} \nonumber\\
\phantom{g_2(x,-13/2)=}{} + \frac{48 (525 x^3 + 840 x^5 + 600 x^7 + 160 x^9 + 16 x^{11})}{1575 + 6300 x^4 + 6720 x^6 + 3600 x^8 + 768 x^{10} + 64 x^{12}},\\
g_3(x,-5/2)= \frac{4 x (27 - 72 x^2 + 16 x^8)}{27 + 54 x^2 + 96 x^6 - 48 x^8 + 32 x^{10}},\\
g_3(x,-9/2)= -\frac{32 (15 x^3 + 12 x^5 + 4 x^7)}{45 + 120 x^4 + 64 x^6 + 16 x^8} \nonumber\\
\phantom{g_3(x,-9/2)=}{} +\frac{24 (225 x - 150 x^3 + 120 x^5 + 240 x^7 + 80 x^9 + 32 x^{11})}{675 + 2700 x^2 - 900 x^4 + 480 x^6 + 720 x^8 + 192 x^{10} + 64 x^{12}}.
\end{gather*}

\section{Conclusions}\label{section8}
In the f\/irst part of this paper we have reviewed the main results concerning the most general Schr\"odinger Hamiltonians characterized by second-order PHA, i.e., possessing third-order dif\/ferential ladder operators. In particular, it was seen that the corresponding potentials can be obtained from the solutions to~$P_{\rm IV}$.

On the other hand, starting from the $k$-th order SUSY partners of the harmonic oscillator potential, a prescription for generating solutions of $P_{\rm IV}$ has been introduced. We have shown that the Hamiltonians associated to these solutions have two independent physical ladders: an inf\/inite one starting from $1/2$ and a f\/inite one placed completely below~$1/2$. We also have identif\/ied three solution hierarchies of the~$P_{\rm IV}$ equation, namely, conf\/luent hypergeometric, error function, and rational hierarchies, as well as some explicit expressions for each of them.

Inside the idea of spectral manipulation, it would be interesting to investigate the possibility of constructing potentials with more freedom for the position of the f\/inite physical ladder, e.g., some or all levels of this ladder could be placed above~$1/2$. This is a subject of further investigation which we hope to address in the near future.

\subsection*{Acknowledgements}

The authors acknowledge the support of Conacyt.

\pdfbookmark[1]{References}{ref}
\LastPageEnding

\end{document}